\documentclass[12pt,preprint]{aastex}
\shorttitle{Rubidium Isotope Ratio}
\shortauthors{Federman, Knauth, \& Lambert}

\begin{document}
\title{The Interstellar Rubidium Isotope Ratio toward 
Rho Ophiuchi A}
\author{S.R. Federman\altaffilmark{1}$^,$\altaffilmark{2}, 
David C. Knauth\altaffilmark{3}, and David L. Lambert\altaffilmark{4}}
\altaffiltext{1}{Department of Physics and Astronomy, University of Toledo, 
Toledo, OH 43606; \linebreak 
sfederm@uoft02.utoledo.edu.}
\altaffiltext{2}{Guest Observer, McDonald Observatory, University of Texas 
at Austin.}
\altaffiltext{3}{Department of Physics and Astronomy, The Johns Hopkins 
University, 3400 North Charles Street, Baltimore, MD 21218; 
dknauth@pha.jhu.edu.}
\altaffiltext{4}{Department of Astronomy, University of Texas, Austin, TX, 
78712; dll@astro.as.utexas.edu.}

\begin{abstract}

The isotope ratio, $^{85}$Rb/$^{87}$Rb, places constraints on models 
of the nucleosynthesis of heavy elements, 
but there is no precise determination of 
the ratio for material beyond the Solar System.  We report the 
first measurement of the interstellar Rb isotope ratio.  Our 
measurement of the Rb~{\small I} line at 7800 \AA\ 
for the diffuse gas toward $\rho$ Oph A yields a 
value of $1.21 \pm 0.30$ (1-$\sigma$) that differs significantly from 
the meteoritic value of 2.59.  The Rb/K elemental abundance ratio for 
the cloud also is lower than that seen in meteorites.  Comparison of 
the $^{85}$Rb/K and $^{87}$Rb/K ratios with meteoritic values 
indicates that the interstellar $^{85}$Rb abundance in this direction 
is lower than the Solar System abundance.  We attribute the lower 
abundance to a reduced contribution from the $r$-process.  
Interstellar abundances for Kr, Cd, and Sn are consistent with much 
less $r$-process synthesis for the solar neighborhood compared to the 
amount inferred for the Solar System.

\end{abstract}


\section{Introduction}

Studies on rubidium in stellar atmospheres and 
the interstellar medium (ISM) are few, but it is not for a lack of 
interest in its abundance.  The production of Rb involves the neutron
capture $s$- and $r$-processes.  Rubidium production by the
$s$-process occurs through the `weak' process in the He- and C-burning
layers of massive stars and through the `main' process in the He-shell
of low-mass AGB stars.  Supernovae from core collapse of massive
stars are the likely source of the $r$-process.  
Analysis of the Solar System abundances provides estimates
of the fractional responsibility of these neutron capture
processes for the Rb isotopes $^{85}$Rb and $^{87}$Rb.  Our estimates 
drawn from averaging the results of 
Beer, Walter, \& K\"{a}ppeler (1992), Raiteri et al. (1993),
and Arlandini et al. (1999) are that $^{85}$Rb is about 35\% $s$- and
65\% $r$-process in origin with the weak $s$-process being one third 
of the $s$-process contribution.  For $^{87}$Rb, the fractions
are about 70\% from the $s$- and 30\% from the $r$-process with
a minor contribution from the weak $s$-process.  As a result of the
different relative contributions of the $s$- and $r$-processes
to the two isotopes, a measurement of the Rb isotope ratio is
a clue to the history of heavy element nucleosynthesis.

Unfortunately, rubidium is a difficult element to measure.  It is
potentially measureable only in cool stars. Resonance lines of
Rb~{\small I} are detectable at 7800 and 7947 \AA, but blending
with atomic and/or molecular lines adds an unwelcome difficulty to an
abundance analysis.  Rubidium abundances for unevolved stars
were reported by Gratton \& Sneden (1994) and Tomkin \&
Lambert (1989).  Measurements of the abundance for giant stars
were obtained by Lambert et al. (1995), Abia \& Wallerstein (1998),
and Abia et al. (2001).  The primary aim of these latter
studies was to exploit the use of Rb as a neutron densitometer
for the $s$-process site (Lambert et al. 1995).  This use
arises from the role of $^{85}$Kr as a branch in the
$s$-process path.  The branch directly affects the isotope
ratio, $^{85}$Rb/$^{87}$Rb, but it is not directly
measureable from even very high resolution stellar spectra (Lambert 
\& Luck 1976) because the lines are too broad.  Earlier interstellar 
measurements yielded upper limits (Federman et al. 1985) because 
the Rb abundance is quite small ($2.5 \times 10^{-10}$ 
in meteorites; Anders \& Grevesse 1989), and it is derived from 
Rb~{\small I}, not the dominant form Rb~{\small II}.  The only 
interstellar detection so far comes from 
observations on two heavily reddened stars in 
Cygnus (Gredel, Black, \& Yan 2001).  [An earlier reported detection 
by Jura \& Smith (1981) could not be confirmed by Federman et al. 
(1985).]  Here, we present high-resolution 
spectra revealing not only another interstellar detection, but also 
the first measurement determining the $^{85}$Rb/$^{87}$Rb 
isotope ratio for extrasolar gas.

\section{Observations and Analysis}

The star $\rho$ Oph A is an ideal target for a study on 
Rb isotopes.  It is relatively bright ($V$ $=$ 5.0), 
moderately reddened with $E$($B-V$) $=$ 0.47, and has one main 
interstellar component (e.g., Lemoine et al. 1993).  The weak 
lines of Li~{\small I} (Lemoine et al. 1993) and K~{\small I} 
$\lambda$4044 (Crutcher 1978) toward this star also show more 
absorption than is typical for a diffuse cloud.  

The data on Rb~{\small I} $\lambda$7800 
were acquired with the 2dcoud\'{e} spectrograph 
(Tull et al. 1995) on the Harlan J. Smith 2.7 m telescope at McDonald 
Observatory in 2003 May.  We used the high-resolution mode with 
echelle grating E2, centered on 7444 \AA\ in order 46.  The spectra 
were imaged onto a $2048 \times 2048$ Tektronics CCD (TK3).  
The 145 $\mu$m slit provided a resolving power of 175,000 as 
determined from widths of lines in the Th-Ar comparison spectrum,
sufficient to discern broadening of 
the interstellar lines.  We obtained calibration exposures for 
dark current the first night of the run, bias correction and flat 
fielding each night, and Th-Ar spectra every 2 to 3 hours during 
the night.  Our primary target, $\rho$ Oph A, was observed for a total 
of 8$^h$, with individual exposures of 30$^m$ per frame; this 
procedure minimized the deleterious effects caused by cosmic rays.  
In addition, the unreddened star $\alpha$ Vir was observed for a total 
of 100$^m$ centered on the slit to check for contamination from weak 
telluric lines and CCD blemishes not removed during the flat-fielding 
process.  No contamination is present.  An additional 60$^m$ was 
utilized to trail $\alpha$ Vir along the slit to act as a stellar 
flat field.  The analysis described below utilized 
spectra of the K~{\small I} line at 4044 \AA\ that were acquired for 
another project (Knauth, Federman, \& Lambert 2004).  While details 
will be presented in Knauth et al. (2004), we note that the 
instrumentation and observing procedure were basically the same as 
those for Rb~{\small I}, except echelle grating E1 was used.  There 
is at most a 0.1 km s$^{-1}$ difference in line velocities between 
the two setups according to the dispersion solutions found from 
Th-Ar spectra.

Standard routines within the IRAF environment were used 
to extract one-dimensional spectra that were Doppler-corrected and 
normalized to unity.  The Rb~{\small I} 
spectrum for the interstellar gas toward $\rho$ Oph A is displayed in 
Fig. 1.  The appearance of two `features' arises from the 62 m\AA\ 
hyperfine splitting in $^{85}$Rb combined with the stronger 
(redder) hyperfine component in $^{87}$Rb.  
The rms deviations in the stellar 
continuum yield a signal-to-noise ratio per pixel 
of 1200 for $\rho$ Oph A; there are 2.9 
pixels per resolution element.

As in our earlier work on the Li isotope ratio acquired at the same 
resolution (Knauth et al. 2000), the Rb~{\small I} and K~{\small I} 
lines were fitted to extract column densities for neutral $^{85}$Rb, 
$^{87}$Rb, and K.  The relevant atomic data used for input 
(Morton 1991, 2000) and the 
resulting equivalent widths ($W_{\lambda}$) 
and column densities ($N$) are given in Table 1.  There is one 
main component at $V_{LSR}$ of 1.9 km s$^{-1}$ (Lemoine et al. 1993; 
Pan et al. 2003), which at ultra-high-resolution is seen as 
two in K~{\small I} $\lambda$7699 spectra (Welty \& Hobbs 2001).  
Another component at 3.5 km s$^{-1}$ appears in 
K~{\small I} $\lambda$7699 (Welty \& Hobbs 2001; Pan et al. 2003), 
having $\approx$ 20\% of the column of the main component.  
This second component is marginally seen (at the 2-$\sigma$ level) 
in our K~{\small I} $\lambda$4044 spectrum, but there is no evidence for 
it in the Rb~{\small I} line.  The uncertainties in column densities 
were inferred from a map of chi-squared confidence levels (Fig. 2).  
While optical depth effects are not important for these very weak 
lines, the profile fitting code (see Knauth et al. 2003) determines 
the $b$-value as well (from the difference between measured line 
width and the instrumental width determined from Th-Ar lines).  
We obtained $b$-values of 1.0 and 0.8 km s$^{-1}$ for Rb~{\small I} 
and K~{\small I}; the slight difference is not significant at our 
spectral resolution.

\section{Results}

The primary result of our study is the determination of a Rb isotope ratio 
for the main cloud toward $\rho$ Oph A.  The best fit (with a reduced 
chi squared of 1.43 for 47 degrees of freedom) yields a 
$^{85}$Rb/$^{87}$Rb ratio of $1.21 \pm 0.30$ -- see top panel 
of Fig. 1.  This differs significantly from 
the meteoritic value of 2.59 (Anders \& Grevesse 1989).  We attempted 
to fit our data with the Solar System ratio as well (bottom panel 
of Fig. 1).  The reduced chi squared is worse (2.08) and an F-test 
shows that there is 50\% confidence in this fit.  Examination of 
the lower panel reveals where the difference lies: Neither the blue 
nor red shoulders of the lines are fit well using a Solar System 
ratio.  Furthermore, the goodness of fit can be judged by 
variations in the residuals (data minus fit) outside and inside the 
absorption profile.  The variation in the lower panel is not as 
consistent across the spectrum.  Other syntheses with the Solar 
System ratio were attempted without success.  
Because both shoulders show more absorption than is predicted 
by this ratio, addition of the redder 3.5 km s$^{-1}$ component would 
not improve the fit.  Broader lines could fill in the shoulders of 
the observed Rb~{\small I} profile, as could a two-component fit 
upon shifting the profile by about 1 resolution 
element.  [A $b$-value as large as 1.8 
km s$^{-1}$ is possible from our $\chi^2$ analysis, though it is 
not consistent with other data (e.g., Welty \& Hobbs 2001).]  
However, the $^{85}$Rb hyperfine components are broadened as well, 
yielding a chi squared that is almost double the $\chi^2$ of 
the best fit in both cases.  
Finally, we note that the lines representing different values of 
$^{85}$Rb/$^{87}$Rb in Fig. 2 favor a ratio between 1.0 and 1.5.  The 
Solar System ratio of 2.6 differs by 2- to 3-$\sigma$ in each 
quantity (column of Rb isotope), consistent with the 4.5-$\sigma$ 
difference seen in the uncertainty listed for the $^{85}$Rb/$^{87}$Rb 
ratio.

The Rb/K elemental ratio provides a means to determine 
for this interstellar cloud whether the $^{85}$Rb abundance is lower 
or the $^{87}$Rb abundance is higher relative to the abundances in 
meteorites.  Since Rb~{\small I} and K~{\small I} are minor interstellar 
species, ionization balance is required to extract an elemental 
abundance.  The elemental ratio does not depend on electron density, and 
we assume that the depletion levels 
for alkalis are the same (see Welty \& Hobbs 2001; Knauth et al. 
2003).  Then for the Rb/K ratio, we have

\begin{equation}
\frac{A_g({\rm Rb})}{A_g({\rm K})} = 
\left[\frac{N({\rm Rb~{\small I}})}{N({\rm K~{\small I}})}\right] 
\left[\frac{G({\rm Rb~{\small I}})}{G({\rm K~{\small I}})}\right]
\left[\frac{\alpha({\rm K~{\small I}})}{\alpha({\rm Rb~{\small I}})}\right],
\end{equation}

\noindent where $G$(X) is the photoionization rate corrected for grain 
attentuation and $\alpha$(X) is the rate coefficient 
for radiative recombination.  Comparison of the 
theoretical calculations for radiative recombination of Rb and K 
(Wane 1985; P\'{e}quignot \& Aldrovandi 1986; Wane \& Aymar 1987) 
reveals that the coefficients are the same to within about 10\%, 
confirming the assumption made in other interstellar studies
(Jura \& Smith 1981; Federman et al. 1985; Gredel et al. 2001).  
The photoionization rates for 
Rb~{\small I} and K~{\small I} are $3.42 \times 10^{-12}$ and 
$8.67 \times 10^{-12}$ s$^{-1}$, respectively.  The cross 
sections for Rb~{\small I} are from the theoretical calculations of 
Weisheit (1972) for 1150 to 1250 \AA, scaled to the measurements of 
Marr \& Creek (1968) at longer wavelengths, while those for K~{\small I} 
are from the measurements of Hudson \& Carter (1965, 1967), 
Marr \& Creek (1968), and Sandner et al. (1981).  Since the ionization 
potentials for Rb~{\small I} and K~{\small I} 
are similar, 4.18 and 4.34 eV, there is 
no appreciable differential attenuation.  We use this fact to account for 
Rb~{\small I} absorption below 1150 \AA; based on the total rate for 
K~{\small I}, we applied a 13\% `correction' to $G$(Rb~{\small I}).  
The result is a Rb/K ratio of $(1.3 \pm 0.3) \times 10^{-3}$ compared to the 
Solar System ratio of $(1.9 \pm 0.2) \times 10^{-3}$ (Anders \& 
Grevesse 1989).

Limits or values of the Rb/K ratio are available from earlier studies 
of interstellar Rb as well.  For the clouds toward 
$o$ Per, $\zeta$ Per, and $\zeta$ Oph, the most conservative limits 
on Rb~{\small I} absorption (Federman et al. 1985) are used with our more 
recent determinations of the K~{\small I} column density from the 
weak line at 4044 \AA\ (Knauth et al. 2000, 2003).  The 3-$\sigma$ 
upper limits for $N$(Rb~{\small I}) are $\le 2.8 \times 10^9$, 
$\le 1.9 \times 10^9$, and $\le 3.7 \times 10^9$ cm$^{-2}$, 
respectively.  The limits on the Rb/K ratio then become 
$\le 1.2 \times 10^{-3}$, $\le 1.0 \times 10^{-3}$, and 
$\le 1.8 \times 10^{-3}$.  The Rb~{\small I} detections 
toward Cyg OB2 Nos. 5 and 12 (Gredel et al. 2001) can be 
combined with results from high-resolution K~{\small I} $\lambda$7699 
spectra (McCall et al. 2002).  For the three main molecular 
components ($+4.0$, $+6.5$, and $+12.3$ km s$^{-1}$), we derive 
values for $N$(K~{\small I}) of $9.4(7.7) \times 10^{11}$, 
$8.1(12) \times 10^{11}$, and $6.9(16) \times 10^{11}$ cm$^{-2}$ 
for the gas toward No. 5 (No. 12) when adopting a $b$-value of 1 
km s$^{-1}$.  The respective Rb/K ratios are $1.4 \times 10^{-3}$ 
and $1.2 \times 10^{-3}$.  (The uncertainties for these ratios are hard 
to quantify because the Rb~{\small I} component structure is not 
known and because the K~{\small I} column densities from $\lambda$7699 
are very susceptible to small changes in adopted $b$-value.)  
These comparisons suggest that the interstellar Rb/K ratio 
may be lower than the meteoritic abundance throughout the 
solar neighborhood.

\section{Discussion}

Insight into the nucleosynthesis of Rb is obtainable from
the ratios $^{85}$Rb/K and $^{87}$Rb/K for the main 1.9 km s$^{-1}$ 
component toward $\rho$ Oph A.  We find 
$^{85}$Rb/K = $(0.73 \pm 0.12) \times 10^{-3}$ and
$^{87}$Rb/K = $(0.60 \pm 0.13) \times 10^{-3}$, 
assuming that $G$(Rb~{\small I}) and 
$\alpha$(Rb~{\small I}) are not dependent on isotope.  The Solar 
System ratios are $^{85}$Rb/K = $(1.37 \pm 0.14) \times 10^{-3}$
and $^{87}$Rb/K = $(0.53 \pm 0.06) \times 10^{-3}$.  
These ratios show that $^{85}$Rb is underabundant in
the gas by a factor of about two, but $^{87}$Rb has about the
Solar System abundance.  In other words, the $^{85}$Rb/$^{87}$Rb 
and Rb/K ratios toward 
$\rho$ Oph A are low because there is less $^{85}$Rb.  
The `missing' $^{85}$Rb is comparable to 
predictions for the $r$-process component in stellar models 
(Beer et al. 1992; Raiteri et al. 1993), $\sim$ 65\%.  Moreover, the 
models (Beer et al. 1992; Raiteri et al. 1993; Arlandini et al. 1999) 
indicate that $^{87}$Rb arises mainly from the $s$-process.  
We are led to believe that the lower interstellar abundance 
for $^{85}$Rb is due to less $r$-process synthesis.

It is widely assumed that the site of the $r$-process
is the very deep interior of a Type II supernova.  Potassium is
also a product of a SN~II; it is synthesized in explosive
oxygen burning (e.g., Clayton 2003).  These different sites and
quite different synthesis mechanisms make it likely
that the $r$-process yield of a Rb isotope relative to
the yield of K can vary; it may depend for
example on the mass, metallicity, and rotation
of a SN~II's progenitor.

The proposal that local interstellar gas may be deficient in
$r$-process products relative to Solar System material is
curiously supported by interstellar abundances of other lightly
depleted or undepleted elements with a $r$-process contribution.  
We emphasize directions probing material toward Sco OB2, which
includes $\rho$ Oph A.  For Solar System material, about 50\% of Kr,
50\% of Cd, and 30\% of Sn are attributable to the $r$-process
(Beer et al. 1992; Raiteri et al. 1993; Arlandini et 
al. 1999).  The interstellar Kr abundance within a few hundred
parsecs of the Sun is constant at 60\% of the solar-wind 
value (Cartledge et al. 2001, 2003).  
Though there are fewer determinations of the
interstellar Cd abundance, it is constant at 80\% of the Solar
System value (Sofia, Meyer, \& Cardelli 1999). Sofia et al.
find a Sn abundance that is essentially solar, 
if not slightly greater than solar, for sight lines showing 
little depletion in other elements.  Arsenic, which is mostly 
$r$-process (Beer et al. 1992; Raiteri et al. 1993) 
and has only 1 isotope, is seen 
currently only in three reddened directions (Cardelli et al. 1993; 
Federman et al. 2003).  It has a higher condensation temperature 
(see Cardelli et al. 1993) and thus it has more depletion, 
especially in reddened directions.  Measurements of 
As~{\small II} in lightly reddened sight lines, where little 
depletion is expected, appear necessary to confirm our hypothesis.

Variations in space and time of the mix of $s$- and $r$-process
products in the Galaxy's interstellar medium should be reflected
in the abundances of the heavy elements in unevolved stars.  
Relative abundances of elements such as Sr, Y, and Zr and
heavier elements such as Ba, Ce, and Eu for dwarfs of the thin disk
do not vary from star to star by more than the measurement
errors (Edvardsson et al. 1993; Chen et al. 2000; Reddy et al. 
2003).  These measurements would seem to 
exclude the larger variations expected
for stars formed from gas with a large amplitude in the ratio of
$s$- to $r$-process products.  Our interstellar results suggest 
that differences in $n$-capture nucleosynthesis may be local to 
the solar neighborhood.

\section{Concluding Remarks}

Our estimate of the Rb isotope and the Rb/K ratios combined 
with the interstellar abundances of Kr, Cd, and Sn raise the
intriguing possibility that the $r$-process products
for such light $n$-capture 
elements are relatively underrepresented in the
local interstellar medium relative to their contribution in
the 4.5 Gyr older Solar System material.  
While models of the $s$-process are becoming quite 
sophisticated and generally give similar results, differences limit 
the ability to make definitive statements.  For these rarer elements, 
the Solar System abundance is sometimes a matter of debate; Kr is 
an example.  The precision of the atomic data is another factor.  For 
Sn~{\small II}, the $f$-values adopted by Sofia et al. (1999) are 
now known to be valid (Schectman et al. 2000; Alonso-Medina, Col\'{o}n, 
\& Mati\'{n}ez 2003).  A key test of our hypothesis is the measurement 
of accurate interstellar abundances for additional elements,
particularly those derived mainly from the $r$-process.

\acknowledgments
We thank Dr. B. McCall, who kindly provided us with his K~{\small I} 
spectra, and Dr. D. York for helpful suggestions.  
This work was supported in part by NASA LTSA grant 
NAG5-4957 to the Univ. of Toledo.

\clearpage

\begin{deluxetable}{lccccc}
%\rotate
\tablecolumns{6}
\tablewidth{0pt}
\tabletypesize{\scriptsize}
\tablecaption{Input Data and Results}
\startdata
\hline \hline\\
Species & $\lambda$ & $F_l \rightarrow F_u$ & $f$-value & $W_{\lambda}$ & 
$N$ \\
 & (\AA) & & & (m\AA) & (cm$^{-2}$) \\ \hline
$^{85}$Rb~{\tiny I} & 7800.232\ $^a$ & $2 \rightarrow 4,3,2,1$\ $^a$ & 
$2.90 \times 10^{-1}$\ $^a$ & $0.40 \pm 0.04(\rm obs) \pm 0.07 (\rm sys)$ & 
$(2.56 \pm 0.38) \times 10^9$ \\
 & 7800.294\ $^a$ & $3 \rightarrow 4,3,2,1$\ $^a$ & 
$4.06 \times 10^{-1}$\ $^a$ & $0.56 \pm 0.04(\rm obs) \pm 0.07 (\rm sys)$ & 
\\
$^{87}$Rb~{\tiny I} & 7800.183\ $^a$ & $1 \rightarrow 3,2,1$\ $^a$ & 
$2.61 \times 10^{-1}$\ $^a$ & $0.30 \pm 0.04(\rm obs) \pm 0.07 (\rm sys)$ & 
$(2.12 \pm 0.43) \times 10^9$ \\
 & 7800.321\ $^a$ & $2 \rightarrow 3,2,1$\ $^a$ & 
$4.35 \times 10^{-1}$\ $^a$ & $0.40 \pm 0.04(\rm obs) \pm 0.07 (\rm sys)$ & 
\\
K~{\tiny I} & 4044.143\ $^b$ & $^c$ & $6.09 \times 10^{-3}$\ $^b$ & 
$1.19 \pm 0.09(\rm obs)$\ $^d$ & $(1.38 \times 0.11) \times 10^{12}$\ $^d$ \\
\enddata
\tablenotetext{a}{Morton 2000.}
\tablenotetext{b}{Morton 1991.}
\tablenotetext{c}{Hyperfine structure not resolved.}
\tablenotetext{d}{For main component at 1.9 km s$^{-1}$.}
\end{deluxetable}

\clearpage

\begin{figure}
\epsscale{0.67}
\plotone{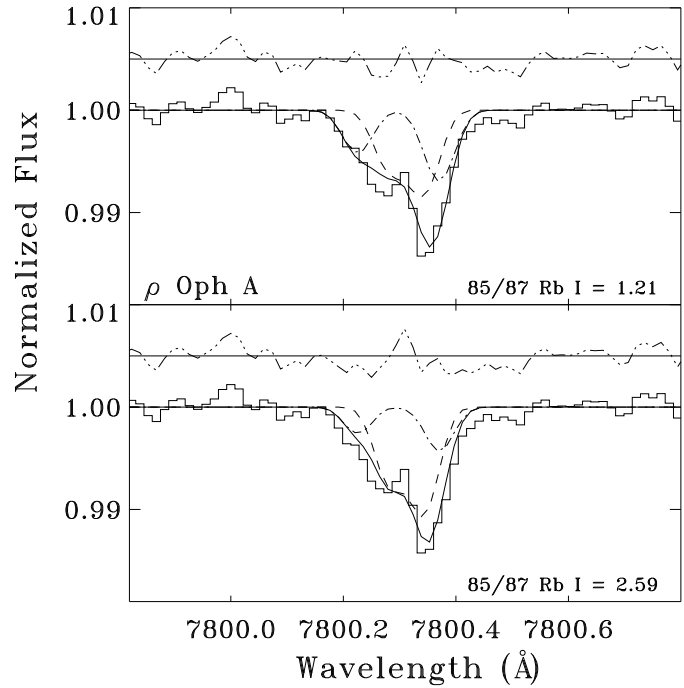}
\vspace{0.3in}
\caption{Rb~{\small I} toward $\rho$ Oph A.  The best fit appears in the 
top panel, and a fit forcing the Rb isotope ratio to be the Solar 
System value is in the bottom panel.  Notice the lack of agreement in 
the shoulders of the line in the lower panel.  The dashed lines show 
the contributions from $^{85}$Rb and the dot-dashed ones from $^{87}$Rb.  
The line above each spectrum indicates data-fit, offset to 1.005.}
\end{figure}

\begin{figure}
\epsscale{0.67}
\plotone{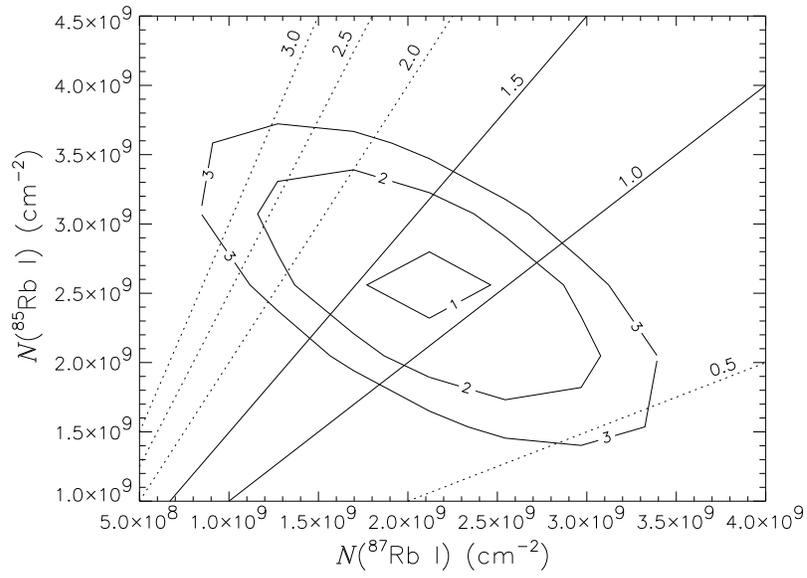}
\caption{A map of chi-squared confidence levels for the best fit to 
Rb~{\small I} absorption.  Contours for 1, 2, and 3 sigma are shown.  
The straight lines shown as overlays represent different 
$^{85}$Rb/$^{87}$Rb ratios.}
\end{figure}

\end{document}